\documentclass[aps,prl,twocolumn,showpacs,superscriptaddress]{revtex4} 
\usepackage{graphicx}
\usepackage{xspace}
\usepackage{siunitx}
\usepackage{multirow}
\usepackage{lineno}
\newcommand{\met}{\ensuremath{\slash \kern-1.5ex E_T}\kern-.35ex\xspace}

\begin{document}
%\pagewiselinenumbers
%\hspace{5.2in} \mbox{Fermilab-XXXXXXX} % ADD IN WITH PREPRINT #

%\linenumbers
%\preprint{}

\title{A search for dark matter in events with one jet and missing transverse energy in $p\bar{p}$ collisions at $\sqrt{s} = 1.96$~TeV}
%\author{1st Draft to Collaboration}
%\date{\today}
\affiliation{Institute of Physics, Academia Sinica, Taipei, Taiwan 11529, Republic of China}
\affiliation{Argonne National Laboratory, Argonne, Illinois 60439, USA}
\affiliation{University of Athens, 157 71 Athens, Greece}
\affiliation{Institut de Fisica d'Altes Energies, ICREA, Universitat Autonoma de Barcelona, E-08193, Bellaterra (Barcelona), Spain}
\affiliation{Baylor University, Waco, Texas 76798, USA}
\affiliation{Istituto Nazionale di Fisica Nucleare Bologna, $^{ee}$University of Bologna, I-40127 Bologna, Italy}
\affiliation{University of California, Davis, Davis, California 95616, USA}
\affiliation{University of California, Los Angeles, Los Angeles, California 90024, USA}
\affiliation{Instituto de Fisica de Cantabria, CSIC-University of Cantabria, 39005 Santander, Spain}
\affiliation{Carnegie Mellon University, Pittsburgh, Pennsylvania 15213, USA}
\affiliation{Enrico Fermi Institute, University of Chicago, Chicago, Illinois 60637, USA}
\affiliation{Comenius University, 842 48 Bratislava, Slovakia; Institute of Experimental Physics, 040 01 Kosice, Slovakia}
\affiliation{Joint Institute for Nuclear Research, RU-141980 Dubna, Russia}
\affiliation{Duke University, Durham, North Carolina 27708, USA}
\affiliation{Fermi National Accelerator Laboratory, Batavia, Illinois 60510, USA}
\affiliation{University of Florida, Gainesville, Florida 32611, USA}
\affiliation{Laboratori Nazionali di Frascati, Istituto Nazionale di Fisica Nucleare, I-00044 Frascati, Italy}
\affiliation{University of Geneva, CH-1211 Geneva 4, Switzerland}
\affiliation{Glasgow University, Glasgow G12 8QQ, United Kingdom}
\affiliation{Harvard University, Cambridge, Massachusetts 02138, USA}
\affiliation{Division of High Energy Physics, Department of Physics, University of Helsinki and Helsinki Institute of Physics, FIN-00014, Helsinki, Finland}
\affiliation{University of Illinois, Urbana, Illinois 61801, USA}
\affiliation{The Johns Hopkins University, Baltimore, Maryland 21218, USA}
\affiliation{Institut f\"{u}r Experimentelle Kernphysik, Karlsruhe Institute of Technology, D-76131 Karlsruhe, Germany}
\affiliation{Center for High Energy Physics: Kyungpook National University, Daegu 702-701, Korea; Seoul National University, Seoul 151-742, Korea; Sungkyunkwan University, Suwon 440-746, Korea; Korea Institute of Science and Technology Information, Daejeon 305-806, Korea; Chonnam National University, Gwangju 500-757, Korea; Chonbuk National University, Jeonju 561-756, Korea}
\affiliation{Ernest Orlando Lawrence Berkeley National Laboratory, Berkeley, California 94720, USA}
\affiliation{University of Liverpool, Liverpool L69 7ZE, United Kingdom}
\affiliation{University College London, London WC1E 6BT, United Kingdom}
\affiliation{Centro de Investigaciones Energeticas Medioambientales y Tecnologicas, E-28040 Madrid, Spain}
\affiliation{Massachusetts Institute of Technology, Cambridge, Massachusetts 02139, USA}
\affiliation{Institute of Particle Physics: McGill University, Montr\'{e}al, Qu\'{e}bec, Canada H3A~2T8; Simon Fraser University, Burnaby, British Columbia, Canada V5A~1S6; University of Toronto, Toronto, Ontario, Canada M5S~1A7; and TRIUMF, Vancouver, British Columbia, Canada V6T~2A3}
\affiliation{University of Michigan, Ann Arbor, Michigan 48109, USA}
\affiliation{Michigan State University, East Lansing, Michigan 48824, USA}
\affiliation{Institution for Theoretical and Experimental Physics, ITEP, Moscow 117259, Russia}
\affiliation{University of New Mexico, Albuquerque, New Mexico 87131, USA}
\affiliation{The Ohio State University, Columbus, Ohio 43210, USA}
\affiliation{Okayama University, Okayama 700-8530, Japan}
\affiliation{Osaka City University, Osaka 588, Japan}
\affiliation{University of Oxford, Oxford OX1 3RH, United Kingdom}
\affiliation{Istituto Nazionale di Fisica Nucleare, Sezione di Padova-Trento, $^{ff}$University of Padova, I-35131 Padova, Italy}
\affiliation{University of Pennsylvania, Philadelphia, Pennsylvania 19104, USA}
\affiliation{Istituto Nazionale di Fisica Nucleare Pisa, $^{gg}$University of Pisa, $^{hh}$University of Siena and $^{ii}$Scuola Normale Superiore, I-56127 Pisa, Italy}
\affiliation{University of Pittsburgh, Pittsburgh, Pennsylvania 15260, USA}
\affiliation{Purdue University, West Lafayette, Indiana 47907, USA}
\affiliation{University of Rochester, Rochester, New York 14627, USA}
\affiliation{The Rockefeller University, New York, New York 10065, USA}
\affiliation{Istituto Nazionale di Fisica Nucleare, Sezione di Roma 1, $^{jj}$Sapienza Universit\`{a} di Roma, I-00185 Roma, Italy}
\affiliation{Rutgers University, Piscataway, New Jersey 08855, USA}
\affiliation{Texas A\&M University, College Station, Texas 77843, USA}
\affiliation{Istituto Nazionale di Fisica Nucleare Trieste/Udine, I-34100 Trieste, $^{kk}$University of Udine, I-33100 Udine, Italy}
\affiliation{University of Tsukuba, Tsukuba, Ibaraki 305, Japan}
\affiliation{Tufts University, Medford, Massachusetts 02155, USA}
\affiliation{University of Virginia, Charlottesville, Virginia 22906, USA}
\affiliation{Waseda University, Tokyo 169, Japan}
\affiliation{Wayne State University, Detroit, Michigan 48201, USA}
\affiliation{University of Wisconsin, Madison, Wisconsin 53706, USA}
\affiliation{Yale University, New Haven, Connecticut 06520, USA}
\affiliation{SLAC National Accelerator Laboratory, Menlo Park, California 94025, USA}

\author{T.~Aaltonen}
\affiliation{Division of High Energy Physics, Department of Physics, University of Helsinki and Helsinki Institute of Physics, FIN-00014, Helsinki, Finland}
\author{B.~\'{A}lvarez~Gonz\'{a}lez$^z$}
\affiliation{Instituto de Fisica de Cantabria, CSIC-University of Cantabria, 39005 Santander, Spain}
\author{S.~Amerio}
\affiliation{Istituto Nazionale di Fisica Nucleare, Sezione di Padova-Trento, $^{ff}$University of Padova, I-35131 Padova, Italy}
\author{D.~Amidei}
\affiliation{University of Michigan, Ann Arbor, Michigan 48109, USA}
\author{A.~Anastassov$^x$}
\affiliation{Fermi National Accelerator Laboratory, Batavia, Illinois 60510, USA}
\author{A.~Annovi}
\affiliation{Laboratori Nazionali di Frascati, Istituto Nazionale di Fisica Nucleare, I-00044 Frascati, Italy}
\author{J.~Antos}
\affiliation{Comenius University, 842 48 Bratislava, Slovakia; Institute of Experimental Physics, 040 01 Kosice, Slovakia}
\author{G.~Apollinari}
\affiliation{Fermi National Accelerator Laboratory, Batavia, Illinois 60510, USA}
\author{J.A.~Appel}
\affiliation{Fermi National Accelerator Laboratory, Batavia, Illinois 60510, USA}
\author{T.~Arisawa}
\affiliation{Waseda University, Tokyo 169, Japan}
% ...

\collaboration{CDF Collaboration\footnote{With visitors from
$^a$Istituto Nazionale di Fisica Nucleare, Sezione di Cagliari, 09042 Monserrato (Cagliari), Italy,
$^b$University of CA Irvine, Irvine, CA 92697, USA,
$^c$University of CA Santa Barbara, Santa Barbara, CA 93106, USA,
$^d$University of CA Santa Cruz, Santa Cruz, CA 95064, USA,
$^e$Institute of Physics, Academy of Sciences of the Czech Republic, Czech Republic,
$^f$CERN, CH-1211 Geneva, Switzerland,
$^g$Cornell University, Ithaca, NY 14853, USA,
$^h$University of Cyprus, Nicosia CY-1678, Cyprus,
$^i$Office of Science, U.S. Department of Energy, Washington, DC 20585, USA,
$^j$University College Dublin, Dublin 4, Ireland,
$^k$ETH, 8092 Zurich, Switzerland,
$^l$University of Fukui, Fukui City, Fukui Prefecture, Japan 910-0017,
$^m$Universidad Iberoamericana, Mexico D.F., Mexico,
$^n$University of Iowa, Iowa City, IA 52242, USA,
$^o$Kinki University, Higashi-Osaka City, Japan 577-8502,
$^p$Kansas State University, Manhattan, KS 66506, USA,
$^q$Ewha Womans University, Seoul, 120-750, Korea,
$^r$University of Manchester, Manchester M13 9PL, United Kingdom,
$^s$Queen Mary, University of London, London, E1 4NS, United Kingdom,
$^t$University of Melbourne, Victoria 3010, Australia,
$^u$Muons, Inc., Batavia, IL 60510, USA,
$^v$Nagasaki Institute of Applied Science, Nagasaki, Japan,
$^w$National Research Nuclear University, Moscow, Russia,
$^x$Northwestern University, Evanston, IL 60208, USA,
$^y$University of Notre Dame, Notre Dame, IN 46556, USA,
$^z$Universidad de Oviedo, E-33007 Oviedo, Spain,
$^{aa}$CNRS-IN2P3, Paris, F-75205 France,
$^{bb}$Texas Tech University, Lubbock, TX 79609, USA,
$^{cc}$Universidad Tecnica Federico Santa Maria, 110v Valparaiso, Chile,
$^{dd}$Yarmouk University, Irbid 211-63, Jordan,
}}
\noaffiliation


%\author{Justin Pilot}
%\email[]{pilot@fnal.gov}
%\thanks{}
%altaffiliation{}
%\affiliation{The Ohio State University, Department of Physics, Columbus, OH 43210, USA}


\newcommand{\ppbar}{$p\overline{p}$}
\newcommand{\unit}[1]{\,\mathrm{#1}}  % to remove, replace "\unit" by "\,\mathrm"
\newcommand{\gevcc}{GeV/c$^2$}
\def\missET {{\not\!\!E_T}}
\def\missETvec {{\not\!\!\vec{E}_T}}
\newcommand{\ttbar}{$t\overline{t}$}

\date{\today}

\begin{abstract}
We present the results of a search for dark matter production in the monojet signature.
We analyze a sample of Tevatron \ppbar\ collisions at $\sqrt{s}$=1.96~TeV corresponding to an integrated luminosity of 6.7~$\unit{fb^{-1}}$ recorded by the CDF II detector.
In events with large missing transverse energy and one energetic jet, we find good agreement between the standard model prediction and the observed data.
We set 90\% confidence level upper limits on the dark matter production rate.
The limits are translated into bounds on nucleon--dark matter scattering rates which are competitive with current direct detection bounds on spin-independent interaction below a dark matter candidate mass of 5~\gevcc, and on spin-dependent interactions up to masses of 200~\gevcc.
\end{abstract}

\pacs{12.60.-i,95.35.+d}

% 12.60.-i = models beyond SM
% 95.35.+d  = dark matter 

%\keywords{}

\maketitle

% INTRO PARAGRAPH

The cosmological abundance of dark matter (DM) is now precisely known through the observation of its gravitational interactions~\cite{Komatsu:2008hk}.  
Yet the nature of DM itself remains a mystery, with many models of physics beyond the standard model (SM) proposing DM candidates.
Perhaps the best motivated DM candidate is a new weakly interacting massive particle (WIMP) with mass of \mbox{O(1 -- 1000)}~\gevcc. 
This class of DM candidates appears in many models of new physics with interactions that allow for DM detection 
through WIMP-nucleon scattering in direct detection experiments~\cite{Gaitskell:2004gd}.  

While there is no conclusive evidence for WIMP-nucleon scattering, several recent direct detection experiments have yielded results suggestive of a low-mass ($\sim$10~\gevcc) WIMP~\cite{DAMALIBRA,COGENT,CRESST2}.
In light of these results, there has been significant interest~\cite{BaiFoxHarnik,Fox:2011pm,Goodman:2010yf,Goodman:2010ku} in the potential of collider searches to either observe the production of DM particles (\mbox{$\chi$}), or to constrain the DM production rate.
The collider mode of production that is expected to yield the most stringent bounds on the DM production rate is the monojet (\mbox{$p\bar{p}\rightarrow \chi \bar{\chi} + $jet}) mode, where the jet has a transverse energy of \mbox{O(100)~$\unit{GeV}$} and originates in initial state radiation.
Previous studies of the monojet signature have been performed~\cite{CDFLED,Aad:2011xw,Chatrchyan:2011nd} in the context of searches for large extra dimensions. 

% Modified based on suggestion from BJK
In this Letter, we present the results of the first direct search for collider production of DM in the monojet mode.
We consider several models of DM production that are relevant for direct detection experiments.
We assume $\chi$ is a Dirac fermion~\cite{MAJORANA}, and that production is mediated by a massive state which couples to DM and SM quarks.
For this analysis, we consider three models of dark matter production which consist of vector ($\mathcal{O}_V$), axial-vector  ($\mathcal{O}_{AV}$), and $t$-channel operators  ($\mathcal{O}_t$) as defined in~\cite{BaiFoxHarnik, Fox:2011pm}.
A universal sum over all quark flavors is assumed for these operators.  
In direct detection experiments, the vector operator leads to spin-independent DM scattering, while the axial vector operator is spin-dependent.  
The $t$-channel operator includes both spin-dependent and spin-independent terms.  
By considering the three types of operators, we are able to constrain both spin-independent and spin-dependent DM-nucleon scattering cross sections.

In direct detection experiments, scattering rates are described by an effective theory containing DM in addition to SM fields.  
As the momentum transfer in DM scattering is far lower than the mass of the particle mediating the interaction, an effective theory provides a valid description.  
In a collider environment, with large momentum transfers, the effective theory approach is not necessarily valid and may change the predicted cross-section and kinematics of the DM model \cite{BaiFoxHarnik, arXiv:1103.0240, Fox:2011pm}.  
We thus consider two possibilities: (1) that these contact interactions are also a good description of collider DM production, and (2) that the production of DM at the Tevatron proceeds through the exchange of a new particle. 
The new mediator particles which lead to the operators $\mathcal{O}_V$, $\mathcal{O}_{AV}$, and $\mathcal{O}_t$, are a heavy vector, axial-vector, and a scalar ``squark'' respectively.
When constraining case 1 we implement models of DM production with very heavy mediators, well above the Tevatron reach (at 10~$\unit{TeV}$), while for case 2 we consider light mediators, within the kinematic reach of the Tevatron (100~\gevcc~for $\mathcal{O}_V$ and $\mathcal{O}_{AV}$ and 400~\gevcc~for $\mathcal{O}_t$).  
% Mod -- end
%%% DATASET & DETECTOR 

We perform the analysis utilizing 6.7~$\unit{fb^{-1}}$ of Tevatron \ppbar\ collisions at $\sqrt{s}$=1.96~TeV recorded by the CDF II detector.
%We perform the analysis utilizing a sample of events recorded by the CDF II detector, and corresponding to 6.7~$\unit{fb^{-1}}$ of Tevatron \ppbar\ collisions at $\sqrt{s}$=1.96~TeV.
The CDF II detector is described in detail elsewhere \cite{detector} and consists of tracking systems immersed in a $1.4$~T magnetic field, surrounded by calorimetery that provides coverage for $|\eta| < 3.6$~\cite{coordinates}.  
A system of drift chambers external to the calorimetery provides muon detection capability for $|\eta| < 1.5$. 

%%% TRIGGER
%CDF II records only those collision events that meet the criteria of an online event-selection (trigger) system.
The DM candidate is expected to interact minimally with the CDF II detector resulting in large missing transverse energy ($\missET$)~\cite{MET}. 
We analyze a sample of events consistent with this characteristic, as collected by a $\missET$ online event-selection (trigger) algorithm which selects collision events with $\missET \ge 40~\unit{GeV}$.   
We find that the trigger has a 90\% selection efficiency for events with $\missET = 60 \ \unit{GeV}$, rising to an efficiency of 95\% for $\missET \ge 70 \ \unit{GeV}$.
%%%%% SELECTION

%We impose additional selections on events selected by the $\missET$ trigger algorithm.
The selected events are required to have been recorded with fully-functioning calorimeter, muon, and tracking systems.  
%, and to be centered within the 
%CDF II detector.  %, and to not contain cosmic ray tracks or calorimeter activity that is not coincidental with $p\bar{p}$ collisions.
We require events to have $\missET > 60 \ \unit{GeV}$.  
We reconstruct jets from calorimeter energy deposits using a cone algorithm~\cite{Blazey:1999tm} with a radius in pseudorapidity-azimuth space of 0.4.    
The jet energies are corrected for variations in detector response and instrumentation, and the extra contribution from additional  $p\bar{p}$ pair interactions in the same event~\cite{Bhatti:2005ai}.
Events are required to have exactly one jet with $E_T \ge 60\unit{GeV}$ and $|\eta| < 1.1$.
%We do not consider jets with $E_T < 20\unit{GeV}$.
To reject events arising from non-collision sources, we require significant track activity within the jet cone.
Events in which the jet does not contain at least one track with a transverse momentum ($p_{T}$) of 
at least $10 \ \unit{GeV/c}$ are rejected.
We reject events in which the jets are reconstructed in a partially instrumented regions of the calorimeter.  
%region of poor or incomplete calorimeter coverage.
%We reject events in which the jet is reconstructed in a region of poor or incomplete calorimeter coverage.
%Events with an imbalance in the distribution of energy between the hadronic and electromagnetic calorimetery are also rejected.
%To remove photons we require that the electromagnetic fraction of the total energy deposited in the calorimeter systems to be between 0.35 and 0.85.
To remove photons we require that the electromagnetic fraction of the total energy deposited in the calorimeter systems to be below 0.85.
Similarly, to remove events with muon bremsstrahlung from cosmic rays or beam-detector interactions~\cite{KaragozUnel:2002mk} we require an electromagnetic fraction of greater than 0.35.
To accommodate extra jets arising from initial state radiation, we retain events with one additional jet with an $E_T$ of less than $30~\unit{GeV}$ and $|\eta| < 2.4$.

%%%% QCD Rejection  + nonCollision rejection
The sample of events that pass the above selections is dominated by background contributions from QCD multijet processes in which one (or more) of the jets is mis-reconstructed.  
Improper reconstruction of a jet produces an event topology in which the $\missETvec$ is aligned with the mis-reconstructed jet.
To reduce this background we require a minimum separation in azimuthal angle of $\Delta \phi (\missETvec,~$jet$) > 0.4$ between the direction of $\missETvec$ and that of any jet with $E_T > 20~\unit{GeV}$ in the event.
We also require a separation of $\Delta \phi (\missETvec,~$jet$) > 2.5$ between the direction of $\missETvec$ and the leading jet.
We achieve further reduction of the multijet background to our search by utilizing an artificial neural network (NN) designed to separate multijet events from electroweak processes.
The NN combines event quantities including the separation in azimuthal angle between jets and $\missETvec$, jet energies, $\missET$, and the number of jets, returning a single numerical value for each event. 
In training, the NN was optimized to isolate simulated $Z$ and $W$ boson events from a sample of data events in which the most energetic jet had $E_T < 60~\unit{GeV}$, or in which there were more than three jets.
%The following quantities are input to the NN:
%minimum $\Delta\Phi(\missETvec,~$any jet$)$,
%the ratio of the total jet $E_T$ to the sum of the jet $E_T$ and $\missET$,
%$\Delta\Phi(\missETvec,~$lead-$E_{T}$ jet$)$,
%$\missET$,
%the $E_{T}$ of the leading jet,
%the number of jets, 
%$\Delta\Phi(\missETvec, $TrkMET$_{10})$,
%Magnitude of the TrkMET$_{10}$,
%$\Delta\Phi($TrkMET$_{10}$, lead-$E_{T}$ jet$)$,
%Event EM fraction,
%and the pseudo-rapidity of the lead-$E_{T}$ jet.
We find that approximately 85\% of multijet events produce a NN value of less than 0.3, and reject these events.

%%%%% ISOLATED TRACK VETO

In the remaining sample of events, we expect significant contributions from $Z$ and $W$ boson processes, in which the $Z$ or $W$ decays leptonically.
We reduce the contribution from these processes by vetoing events which contain one or more tracks with $p_{T}  \ge 10 \ \unit{GeV/c}$ that are not embedded within a jet.
%We refer to this selection as the isolated track veto.

%%%%%%% DATA MODEL (MC)
The events passing the above selections form our analysis sample, and are examined for the presence of events arising from DM production.
Within this sample we expect significant contributions from $Z$ boson processes in which the $Z$ boson decays invisibly to neutrinos.  
In addition, we expect $W$ boson processes to contribute whenever the lepton from the leptonically decaying $W$ boson is outside of the acceptance of the CDF tracking system.
We model $W$ and $Z$ boson contributions to our analysis sample using simulated events generated by {\sc Alpgen}~\cite{Mangano:2002ea} with  {\sc Pythia}~\cite{PYTHIA} for particle showering and hadronization.
The $Z$ and $W$ boson background contributions are determined assuming NNLO calculations~\cite{Hamberg:1990np} of the inclusive production rates.

Minor backgrounds include $t\bar{t}$ modeled with {\sc Pythia}, and single-top processes modeled with {\sc MadGraph}~\cite{MAD} plus {\sc Pythia} for particle showering and hadronization.
%We expect minor contamination from $t\bar{t}$ and single-top processes and model the contribution using {\sc Pythia}, and {\sc MadGraph}~\cite{MAD} plus {\sc Pythia} for particle showering and hadronization, respectively.  
A top-quark mass of 172.5~\gevcc~is assumed for the \ttbar~\cite{Moch:2008ai,Martin:2007bv} $s$-channel, and $t$-channel~\cite{Kidonakis:2006bu} processes with cross sections of cross sections of 7.04, 1.05, and 2.1~$\unit{pb}$, respectively. 
We account for the expected diboson ($WW$,$WZ$,$ZZ$)~\cite{CampbellEllis} contributions to our selected sample with a {\sc Pythia} simulation, and normalize 
the rates of the $WW$, $WZ$, and $ZZ$  processes to 11.34, 3.47 and 3.62~pb.
All simulated samples in this analysis include a detailed {\sc Geant}-based detector simulation~\cite{GEANT} and assume CTEQ5L~\cite{CTEQ5L} parton distribution functions.

%%%%%%% DATA MODEL (MultiJet)
%%%%%%% DATA MODEL (MultiJet)
While our NN requirement rejects the main multijet contamination, we model the remaining multijet background using reweighted data events.
%As noted earlier, events originating in multijet processes enter the analysis sample primarily due to mis-reconstruction or mis-measurement of the energy of one or more jets.  
%This renders simulation of the multijet contribution to our selected samples impractical, and reweighted data are the best alternative.  
We determine the likelihood of an event in our analysis sample to have originated from a multijet background process by utilizing a sample of events with relaxed kinematic selections such that events with any number  of jets with $E_T$ greater than 35~$\unit{GeV}$ are accepted.
Events meeting this relaxed selection constitute the derivation sample.
%We refer to these events as the derivation sample.
To maintain exclusivity, we remove all events entering the analysis sample from the derivation sample.
In addition, we denote the probability that a given event originated in a multijet process as the multijet probability (MJP).

The MJP is obtained as the fraction of events in the derivation sample remaining after subtraction of all simulated backgrounds.  We parameterize the MJP using six observables in order to mimic the topological characteristics of multijet events.
These are $\missET$, the number of jets, the minimum separation in azimuthal angle between $\missETvec$ and a jet, the ratio of 
the scalar sum of jet $E_{T}$ to its sum with the $\missET$, the magnitude of the momentum imbalance from tracks with $p_{T}  \ge 10 \ \unit{GeV/c}$, and the $\missET$ significance~\cite{METsig}.

A given event in the analysis sample is assigned a weight by the MJP, as determined by its values of the six observables.
The multijet background is modeled as the weighted sum of all events contributing to the analysis sample.
We find that the above method accurately determines the shape of the multijet background in all observables of interest.
To obtain an appropriate normalization for the multijet contribution, we require that the sum of the number of events predicted by simulation and by the multijet prediction equal the number of data events observed in the sideband region which is defined such that the NN value is between 0.2 and 0.3.

\begin{figure*}[t]
\begin{center}
\includegraphics[height=45mm,width=2\columnwidth]{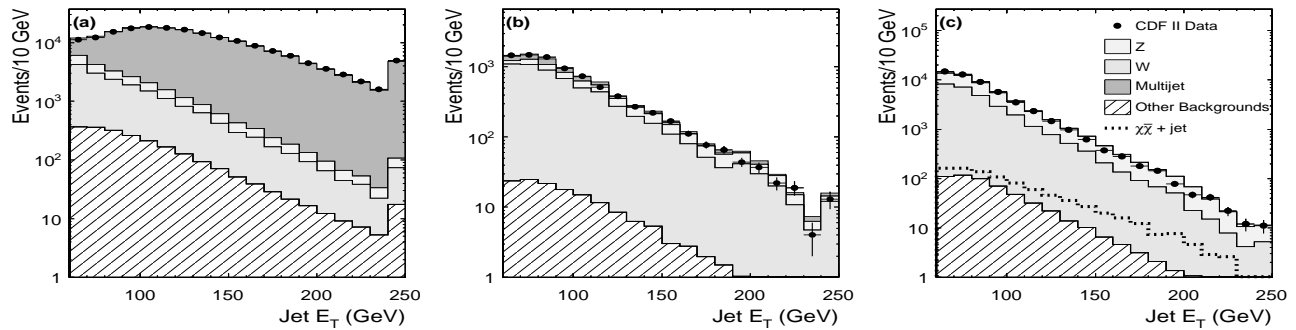}
\caption{Distribution of the jet $E_{T}$ for the multijet control (a), electroweak control (b), and analysis (c) samples.
The last bin contains the overflow.
For the analysis sample, the jet $E_{T}$ of a representative signal process ($\chi\bar{\chi}+$jet) is shown normalized to the 90\% confidence level upper limit production rate of 5.9~pb.}
\label{fig:Et}
\end{center}
\end{figure*}

To test the performance of our data model, we form two additional samples that are exclusive of the analysis sample.
We define an electroweak sample that is composed of events that pass all analysis selection criteria but have one or more tracks with $p_{T}  \ge 10 \ \unit{GeV/c}$, that are not embedded within a jet.
In addition, we define a multijet sample of events passing all the analysis selection requirements except that they have a NN value less than 0.3, $\missETvec$ aligned with a jet, or have more than 2 jets.
We find good agreement between the data and the SM prediction in both control samples. 
The $E_T$ distribution of the leading jet is displayed in Fig.~\ref{fig:Et}.

%\begin{figure}
%\includegraphics[width=1.0\columnwidth]{log_jet1Et.eps}
%\caption{Distribution of the jet $E_{T}$ for the multijet control (left), electroweak control (center), and analysis (right) samples.
%The last bin contains the overflow.
%The jet $E_{T}$ for a representative signal process ($\chi\bar{\chi}+$jet) is shown normalized to the 90\% confidence level upper limit production rate of 5.9~pb.}
%\label{limit_fig}
%\end{figure}

%DM SIGNAL SAMPLES
We model the potential contribution to our analysis sample from a DM signal of \mbox{$p\bar{p}\rightarrow \chi \bar{\chi} + $jet} using a {\sc MadGraph}~\cite{MAD} generator that is interfaced with {\sc Pythia} for showering and hadronization.
We generate variants of the signal models, discussed previously, assuming DM masses between 1 and 300~\gevcc.
We find an efficiency of approximately 2\% when imposing the analysis sample on simulated DM events.

%SYSTEMATICS
Systematic uncertainties affecting the normalization of simulated background components and DM signal arise due to uncertainty in the integrated luminosity ($\sim$6\%), measured jet energy scale ($\sim$7\%), parton distribution function uncertainties ($\sim$2\%), efficiency of the trigger used for data collection ($\sim$2\%), choice of the renormalization scale ($\sim$2\%), and the amount of initial and final state radiation ($\sim$1\%). 
In addition, a 50\% uncertainty is placed on the normalization of the multi-jet prediction.
%, as well as a 100\% uncertainty on the normalization of the non-collision contamination rate. 
The normalization uncertainties for the top~\cite{Moch:2008ai,Martin:2007bv,Kidonakis:2006bu} , $Z/W$~\cite{Abulencia:2005ix}, and diboson~\cite{CampbellEllis} processes
are 10\%, 8\%, and 6\% respectively.
%We assume 10\%, 8\%, and 6\% uncertainties on the normalization of simulated top~\cite{Moch:2008ai,Martin:2007bv,Kidonakis:2006bu} , Z/W~\cite{Abulencia:2005ix}, and diboson~\cite{CampbellEllis} processes respectively.
We include the effect on the jet energy scale uncertainty on the shape of observed quantities, and find this to be the dominant uncertainty.
%We include an uncertainty on the contribution of the multijet background to the shape of the distribution of observed quantities.
%This uncertainty is derived from the observed difference in the multijet shape between the analysis and electroweak samples.
%We include an uncertainty on the distribution shapes of the multijet background, derived from the observed difference in the multijet shapes between the analysis and electroweak samples.
We include an uncertainty on the shape of the multijet background, based on the observed variation in the multijet prediction between the analysis and electroweak samples.

% LIMIT SETTING
The total numbers of observed and expected events in the control and analysis samples is listed in Table~\ref{tab:TOT}.  
In the analysis sample, we observe 52633 events which agrees well with the expectation of 53906	$\pm$	6022. 
As we do not observe a significant excess over the number of events predicted by our background model, we proceed to quantify the maximum allowed DM production cross section.

\begin{figure*}[t]
\begin{center}
\includegraphics[height=45mm,width=1\columnwidth]{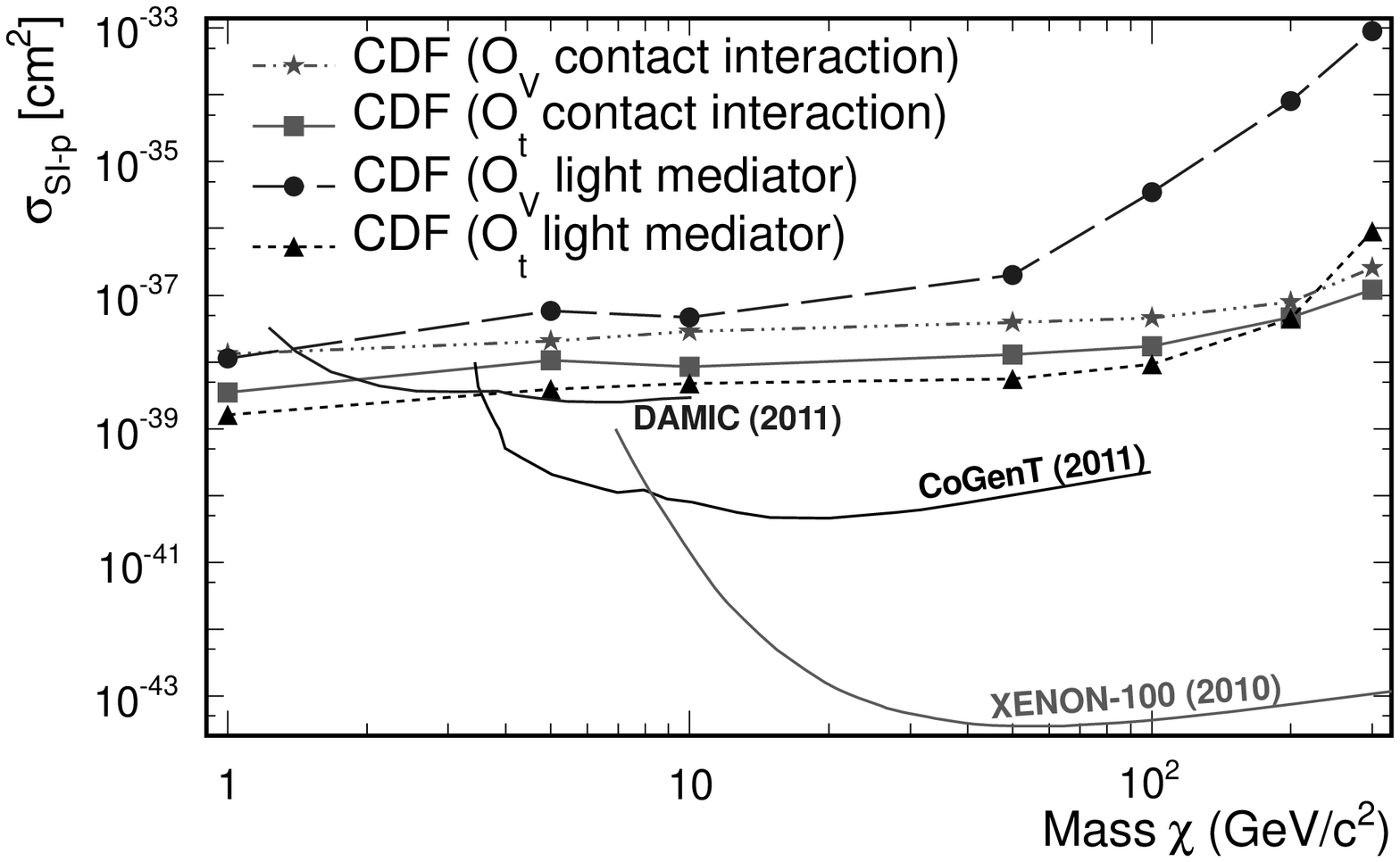}\includegraphics[height=45mm,width=1\columnwidth]{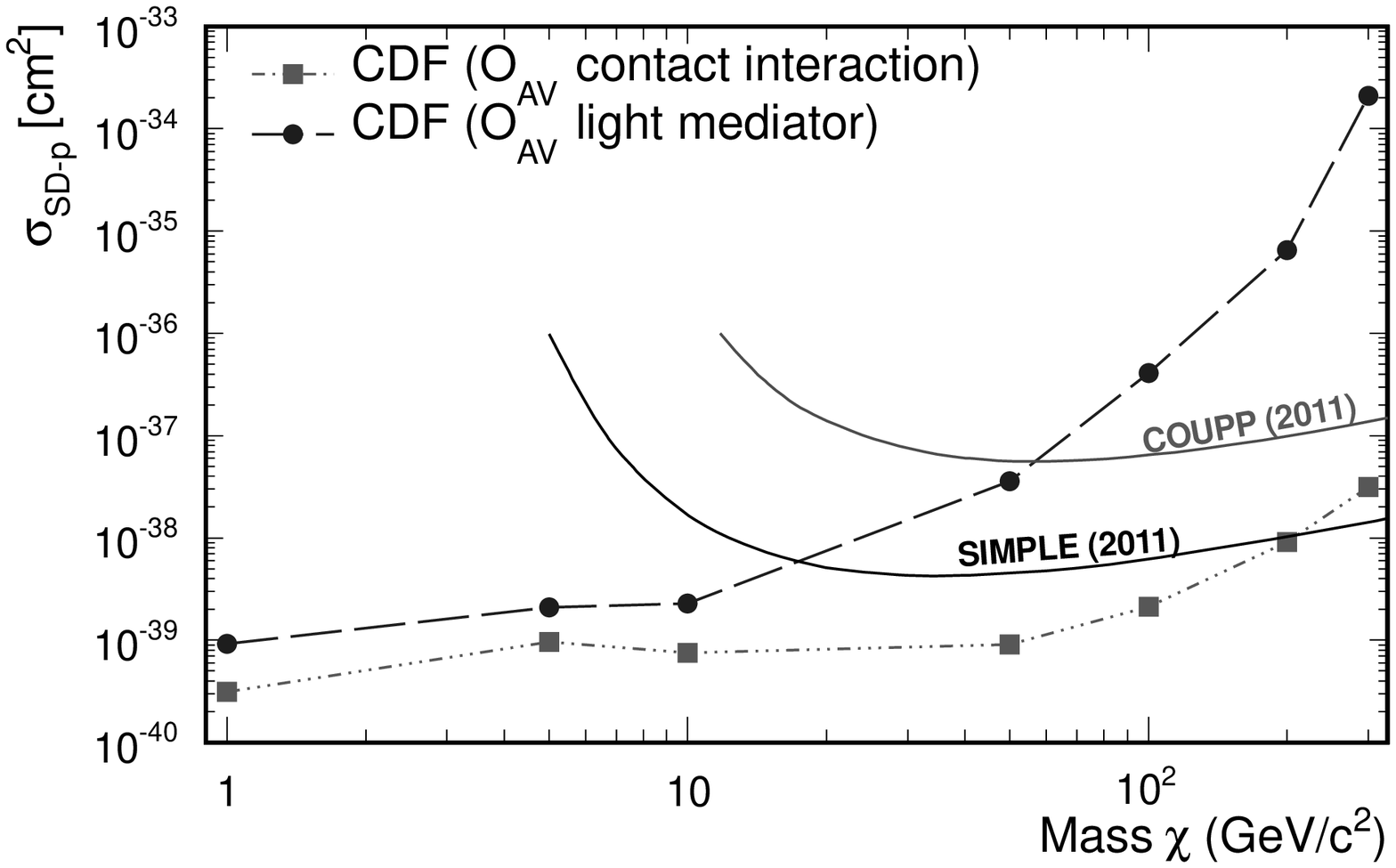}
 \caption{Comparison of CDF results to recent results from DAMIC~\cite{DAMICN}, CoGeNT~\cite{COGENT}, XENON-100~\cite{XENONN}, SIMPLE~\cite{SIMPLEN}, and COUPP~\cite{COUPPN}.
Spin-independent (left) and spin-dependent (right) bounds are shown for the operators (defined in text) $\mathcal{O}_{AV}$, $\mathcal{O}_{V}$, and $\mathcal{O}_{t}$, assuming 
contact interactions.  For comparison we also display CDF bounds assuming light mediators.   }
\label{fig:LIM}
\end{center}
\end{figure*}

We set limits on the DM production rate using a Bayesian likelihood~\cite{PDG2010} formed as a product of likelihoods over bins in the analysis region of the jet $E_T$ distribution. 
We assume a flat prior on the signal rate, and a Gaussian prior for each systematic uncertainty including those affecting sample normalizations and shapes.
We set Bayesian 90\% confidence level upper limits on $\sigma(p\bar{p}\rightarrow \chi \bar{\chi} + \mathrm{jet})$ for each of the models considered.
The expected upper limits at each model point are derived by randomly generating a series of pseudo-datasets, derived from the background prediction, and computing the median of the distribution of resulting upper limits.
The upper limits are listed in Table~\ref{table:lim1}.
We proceed to convert the limits into constraints on the DM-nucleon cross section following~\cite{BaiFoxHarnik,Goodman:2010yf}.
A comparison of the CDF limits to several direct detection results is shown in Fig.~\ref{fig:LIM}.
The CDF limits assuming light mediators are also shown.
The CDF bounds extend beyond the experimental reach of direct detection searches, which are insensitive to DM with a mass of approximately 1~\gevcc.
For a DM mass of 5~\gevcc, CDF bounds on spin-independent interactions are \mbox{O($10^{-38}$)}~$\unit{cm^{2}}$ and are similar to the limits reported by the DAMIC~\cite{DAMICN} collaboration.
In the case of spin-dependent interactions, we report stronger bounds of \mbox{O($10^{-40}$)}~$\unit{cm^{2}}$ for a DM mass of 1~\gevcc, rising to \mbox{O($10^{-39}$)}~$\unit{cm^{2}}$ 
for a mass of 200~\gevcc.
% CONCLUSION

In conclusion, we have performed the first collider search for DM in the monojet production mode.
We have set limits on the DM production rate, and have constrained the spin-independent nucleon-DM scattering rate for a DM mass of roughly 1~\gevcc, and between 1 and 200~\gevcc~for spin-dependent interactions. 
% which translate into bounds on nucleon-dark matter scattering rates.
%The CDF bounds are competitive with current direct detection bounds on spin-independent interactions below a dark matter mass of 5~GeV, 
%and spin-dependent interactions up to masses of 200~GeV.

% SIGNAL REGION TOTALS
\begin{table}
  \caption{Event totals in the analysis and control samples.  Uncertainties are systematic only.}
    \begin{tabular}{c c c c}
     \hline
     \hline
     Source & Multijet  & Electroweak  & Analysis \\
     \hline
$Z$		&	6949	$\pm$	840	&	1280	$\pm$	155	&	22191	$\pm$	2681	\\
$W$		&	14986	$\pm$	2007	&	5582	$\pm$	747	&	27892	$\pm$	3735	\\
Multijet		&	165479	$\pm$	82740	&	1066	$\pm$	533	&	3278	$\pm$	1639	\\
%non-collision		&	49	$\pm$	48	&	1	$\pm$	1	&	6	$\pm$	6	\\
%diboson		&	626	$\pm$	55	&	101	$\pm$	9	&	412	$\pm$	36	\\
%$t\bar{t}$		&	1122	$\pm$	215	&	20	$\pm$	4	&	23	$\pm$	4	\\
%single-top	&	397	$\pm$	54	&	27	$\pm$	4	&	104	$\pm$	14	\\
Other & 2194 $\pm$	233.4 &	149~$\pm$	10.7 &	545~$\pm$	39.3\\
\hline
Total model	&	189608	$\pm$	82787	&	8076	$\pm$	1011	&	53906	$\pm$	6022	\\
\hline
Data	& 188361	&	7942	&	52633	\\							
          \hline
        \hline
         \end{tabular}
   \label{tab:TOT}
\end{table}

\begin{table}[htdp]	
\renewcommand\arraystretch{1.2}
\caption{Expected (Exp.) and observed (Obs.) $90\%$ C.L. upper limits (in pb) on the cross section of 
$p\bar{p}\rightarrow \chi \bar{\chi} + \mathrm{jet}$ for the three operators (defined in text) $\mathcal{O}_{AV}$, $\mathcal{O}_{V}$, and $\mathcal{O}_{t}$, assuming 
contact interactions.  The $\pm1\sigma$ variations on the expected limits are also shown.}
\begin{center}											
\begin{tabular}{c c c c c c c }										
 \hline											
& \multicolumn{2}{c}{---~$\mathcal{O}_{AV}$~---} & \multicolumn{2}{c}{---~$\mathcal{O}_{V}$~---} & \multicolumn{2}{c}{---~$\mathcal{O}_t$~---} \\
 $M_\chi$ & Obs.  & Exp.  & Obs.  & Exp. & Obs.  & Exp. \\
 ($\unit{GeV}/c^2$) & (pb)  & (pb)   & (pb)   &(pb)  &(pb)   & (pb)  \\
 \hline											
\hline
1 &	5.9 & $7.6^{+4.9}_{-3.6}$ &	9.1 & $7.8^{+5}_{-3.5}$ &	6.5 & $7.2^{+4.3}_{-3.6}$ \\
5 &	6.9 & $7.9^{+4.8}_{-3.6}$ &	5.4 & $8.1^{+4.9}_{-4.1}$ &	7.5 & $8.1^{+4.4}_{-3.6}$ \\
10 &	4.5 & $7.9^{+4.8}_{-3.5}$ &	6.4 & $7.7^{+4.4}_{-3.6}$ &	5.0 & $7.0^{+4.2}_{-3.1}$ \\
50 &	3.4 & $7.0^{+4.2}_{-3.4}$ &	6.6 & $7.5^{+4.6}_{-3.4}$ &	5.3 & $6.5^{+4.0}_{-3.1}$ \\
100 &	4.5 & $6.0^{+3.6}_{-2.9}$ &	5.7 & $6.2^{+3.7}_{-2.8}$ &	4.6 & $6.1^{+3.7}_{-2.9}$ \\
200 &	4.8 & $5.6^{+3.2}_{-2.7}$ &	3.9 & $5.6^{+3.6}_{-2.7}$ &	4.2 & $4.8^{+3.2}_{-2.2}$ \\
300 &	3.1 & $6.1^{+3.9}_{-2.7}$ &	3.5 & $5.6^{+3.4}_{-2.4}$ &	2.7 & $5.1^{+3.2}_{-2.4}$ \\
\hline		
 \hline											
  \end{tabular}
 \end{center}											
\label{table:lim1}%\vspace{1cm}
\end{table}

%Acknowledgments
%
We thank the Fermilab staff and the technical staffs of the participating institutions for their vital contributions. This work was supported by the U.S. Department of Energy and National Science Foundation; the Italian Istituto Nazionale di Fisica Nucleare; the Ministry of Education, Culture, Sports, Science and Technology of Japan; the Natural Sciences and Engineering Research Council of Canada; the National Science Council of the Republic of China; the Swiss National Science Foundation; the A.P. Sloan Foundation; the Bundesministerium f\"ur Bildung und Forschung, Germany; the Korean World Class University Program, the National Research Foundation of Korea; the Science and Technology Facilities Council and the Royal Society, UK; the Russian Foundation for Basic Research; the Ministerio de Ciencia e Innovaci\'{o}n, and Programa Consolider-Ingenio 2010, Spain; the Slovak R\&D Agency; the Academy of Finland; and the Australian Research Council (ARC).

\bibliographystyle{apsrev}
\bibliography{mono}

\begin{thebibliography}{37}
\bibitem[{MAJ()}]{MAJORANA}
\bibinfo{note}{Although not considered here we expect similar results for
  Majorana DM, with the exception being that there are no vector interactions
  for Majorana DM.}
\bibitem[{coo()}]{coordinates}
\bibinfo{note}{We use a cylindrical coordinate system with $z$ along the proton
  beam direction, $r$ the perpendicular radius from the central axis of the
  detector, and $\phi$ the azimuthal angle. For $\theta$ the polar angle from
  the proton beam, we define $\eta = -\ln\tan(\theta/2)$, %We denote the
  transverse momentum $p_T = p \sin\theta$ and transverse energy $E_T = E
  \sin\theta$.}
\bibitem[{MET({\natexlab{a}})}]{MET}
\bibinfo{note}{The missing $E_T$ ($\not\!\!\vec{E}_T$) is defined by the sum
  over calorimeter towers: $\not\!\!\vec{E}_T = - \sum_{i} E_T^i \hat{n}_i$,
  where $i =$ calorimeter tower number with $|\eta| < 3.6$, $\hat{n}_i$ is a
  unit vector perpendicular to the beam axis and pointing at the i$^{th}$
  calorimeter tower. We also define $\not\!\!E_T=|{\not\!\!\vec{E}_T}|$.}
\bibitem[{MET({\natexlab{b}})}]{METsig}
\bibinfo{note}{We define the $\missET$ significance as $\frac{\missET}{\sum_{i}
  E_T^i}$ where $i$ runs over all jets with $E_T > 20 \ \unit{GeV}$ and
  detector $|\eta| \leq 2.4$.}
\end{thebibliography}

\end{document}